\documentclass[aps,prd,twocolumn,superscriptaddress,nofootinbib,eqsecnum,floatfix]{revtex4}

\pdfoutput=1

\usepackage{amsfonts}
\usepackage{amsmath}
\usepackage{amssymb}
\usepackage{graphicx,color}
\usepackage{float}
\usepackage{hyperref}
\usepackage{subfigure}
\usepackage{dcolumn}

\usepackage{verbatim}

\begin{document}

\title{Kinetic Equilibration after Preheating}

\author{Robert Brandenberger}
\email{rhb@physics.mcgill.ca}
\affiliation{Department of Physics, McGill University, Montr\'{e}al,
  QC, H3A 2T8, Canada}

\author{Ryo Namba}
\email{namba@physics.mcgill.ca}
\affiliation{Department of Physics, McGill University, Montr\'{e}al,
  QC, H3A 2T8, Canada}

\author{Rudnei O. Ramos} \email{rudnei@uerj.br}
\affiliation{Departamento de Fisica Teorica, Universidade do Estado do
  Rio de Janeiro,  20550-013 Rio de Janeiro, RJ, Brazil}

\begin{abstract}

We study thermal equilibration after preheating in inflationary
cosmology, which is an important step  towards a comprehensive
understanding of cosmic thermal history.  By noticing that the problem
is  parallel to  thermalization after a relativistic heavy ion
collision, we  make use of the methods developed in this context and
that seek for an analytical approach to the Boltzmann equation.  In
particular, an exact solution for number-conserving scatterings is
available for the distribution function in a
Friedmann-Lema\^{i}tre-Robertson-Walker metric and can be utilized for
the spectral evolution of kinetic equilibration  process after
preheating. We find that thermal equilibration is almost instantaneous
on the time scale of the Hubble time.  We also make an explicit
prediction for the duration (the number of e-folds of expansion)
required for this process of thermal equilibration to complete
following the end of inflation.

\end{abstract}

\maketitle

\section{Introduction} 
\label{sec:intro}

The inflationary scenario~\cite{Guth} has become the main paradigm of
early universe cosmology. It involves a phase of accelerated expansion
of space which is usually driven by the potential energy of a scalar
matter field $\varphi$, the so-called {\it inflaton}. The phase of
accelerated expansion redshifts whatever regular matter was present at
the beginning of the period of inflation and leaves the matter fields
in a vacuum state.%
\footnote{One should note that one exception to this
is the {\it warm inflation} scenario~\cite{Berera:2008ar} in which
there is a continuous and sustained production of radiation  during
inflation.} 
In order to make successful contact with late-time
cosmology, a phase of {\it reheating} during which the energy is
transferred from the inflaton field to regular matter,
in the form of relativistic particles, is hence essential.

Reheating was initially studied in leading-order perturbation
theory~\cite{DL, AFW}. Such analysis, however, neglect the fact that
the inflaton is a coherent state of zero momentum particles at the end
of the period of inflation. In many models, the inflaton field is
slowly rolling during the period of accelerated expansion but then
begins to oscillate about the minimum of its potential energy function
when inflation ends. As realized in Refs.~\cite{TB, DK},  these
oscillations lead to a parametric instability in the equations of
motion for modes of fields $\chi$ which couple to the inflaton. There
is a {}Floquet instability~\cite{Floquet}  that can then lead to an
energy transfer from the inflaton to matter which is rapid on the
scale of the Hubble time. However, the matter that is produced by this
process is in general not thermally distributed. There are models in which the
resonance occurs in narrow instability bands, but there are also
classes of models in which {\it broad resonance} occurs, a process in
which all modes of $\chi$ with wavenumbers smaller than a critical
value $k_c$, set by the mass of $\chi$, are excited. This process has
been dubbed {\it preheating} in the literature~\cite{KLS,STB,KLS2}
(see also Refs.~\cite{ABCM,Karouby} for recent detailed reviews of
preheating). 

In order to connect successfully to late-time cosmology, the
preheating phase must be followed by a period of thermalization. There
are two aspects to the thermalization process. The first involves the
redistribution of the energy of the particles such that a thermal
distribution is obtained. This is called {\it kinetic
  equilibration}. The second aspect is the increase in the number of
particles required in order to obtain the thermal amplitude of the
particle spectrum. This is called {\it chemical equilibration}.  In
this work we will focus only on kinetic equilibration. Thermalization
after inflation has been studied in many works (see, e.g.,
Refs.~\cite{ABCM, Karouby} and the cited references therein and also
Ref.~\cite{original} for some early studies), but mostly using a
leading-order perturbative analysis. Using these methods it is usually
found that the time it takes before thermalization completes is much
longer than the Hubble expansion time at the end of inflation, and
hence, that the highest temperature of the post-inflationary thermal
gas can be in general many orders of magnitude smaller than that
corresponding to the energy scale of inflation. 

However, since the state of matter after preheating is a special state
which is far from thermal equilibrium, the applicability of a
leading-order perturbative analysis is questionable. In the present
work we wish to present an analysis of kinetic equilibration, the
first stage of thermalization, which is appropriate for initial states
of matter produced by preheating. The problem of thermalization has
also been considered using numerical simulations of classical field
dynamics~\cite{numerical}, but for initial conditions which follow a
period of narrow resonance, not broad resonance (see also
Ref.~\cite{Berges}). 

We consider models in which the initial energy transfer from the
inflaton field to regular matter occurs via broad parametric
resonance. In this case, the state at the end of preheating is similar
to the initial state assumed in some studies of thermalization in
heavy ion collisions (see, e.g., Ref.~\cite{RHIC}). There has recently
been a lot of work on thermalization of plasmas created in
relativistic heavy ion collisions. Here we take advantage of this
similarity between preheating and the dynamics that can happen in the
early stages of thermalization following heavy ion collision
experiments and apply some of those techniques used to study the
latter in the present study.  We in particular make use of the
formalism recently developed in Ref.~\cite{Denicol}, which we find
well suited to the present problem.

This work is organized as follows. In Sec.~\ref{sec:kinetic}, we
briefly review some of the key aspects of the preheating dynamics.  In
Sec.~\ref{sec3}, we summarize the formalism developed in
Ref.~\cite{Denicol} and that we make use of.  Then, in
Sec.~\ref{sec4}, we specifically apply the results shown in the
previous section to the problem of kinetic equilibration of the
produced particles following preheating. In particular, we
compute the time scale over which the distribution of particles
initially created by broad resonance preheating approaches a thermal
distribution. We end with a discussion of our results in
Sec.~\ref{conclusions}.

We work in natural units, in which $\hbar = c = k_B = 1$. The reduced
Planck mass is denoted by $M_{\rm Pl} \equiv
1/\sqrt{8 \pi G} \simeq 2.4 \times 10^{18}$GeV. We assume a spatially
flat Friedmann-Lema\^{i}tre-Robertson-Walker (FLRW) metric,
\begin{equation}
ds^2 = - dt^2 + a(t)^2 d{\bf x}^2  ,
\end{equation}
where $t$ is the physical time, ${\bf x}$ are the comoving spatial
coordinates and $a(t)$ is the scale factor, which describes the
expansion of space. The Hubble expansion rate is
\begin{equation}
H(t) \equiv \frac{{\dot a}}{a} ,
\end{equation}
where the overdot is the derivative with respect to $t$. The inverse
of $H(t)$ is the Hubble expansion time, the characteristic expansion
time scale. Preheating proceeds on a time scale much shorter than
$H^{-1}(t)$, whereas leading-order perturbative reheating calculations
lead to a reheating time that is much longer than $H^{-1}(t)$. 
Our interest is to compare the time of kinetic equilibration with this
Hubble expansion time.
 
\section{A brief review of preheating} 
\label{sec:kinetic}

In this work we mostly consider only one particle species
that is involved in the thermalization process, which is the species
corresponding to the field which is excited during broad resonance
preheating. However, we also comment later on the effects of
having more than one particle species and that could contribute to the
radiation bath (i.e., relativistic species). Thus, we consider a
theory in which the inflaton field $\varphi$ couples to another scalar
field $\chi$ via an interaction Lagrangian density given by
\begin{equation}
{\cal L}_I  =  - \frac{g^2}{2} \, \chi^2 \varphi^2 \, ,
\label{L_I}
\end{equation}
where $g$ is the coupling constant between $\varphi$ and $\chi$.  We
have in mind a model of {\it chaotic inflation}~\cite{Linde}, where
$\varphi$ begins to oscillate about $\varphi = 0$ at the end of the
inflationary phase. The amplitude ${\cal A}$ of oscillation is close
to the Planck mass (about one order of magnitude smaller), while the
frequency is given by the mass of $\varphi$, $m_\varphi$, which must be
several orders of magnitude smaller than the Planck mass,
$m_\varphi\sim 10^{-6} M_{\rm Pl}$, in order to obtain the correct
order of magnitude of the cosmological perturbations (see, e.g.,
Refs.~\cite{MFB, RHBfluctRev} for reviews of the theory of
cosmological perturbations). 

In the presence of the oscillating inflaton field, the equation of
motion for the Fourier modes $\chi_k$ of $\chi$ is
\begin{equation}
\chi_k^{\prime \prime} + \left[A_k - 2 q \cos(2z) \right] \chi_k  =  0
,
\end{equation}
where $z = m_\varphi t$, a prime denotes the derivative with respect
to $z$ and the parameters $q$ and $A_k$ are given, respectively by
\begin{eqnarray}
q  &=&  \frac{g^2 {\cal A}^2}{4 m_\varphi^2} ,
\label{eqq} \\
A_k  &=&  \frac{k^2 + m_{\chi}^2}{m_\varphi^2} + 2 q ,
\label{eqAk}
\end{eqnarray}
where $m_{\chi}$ is the mass of $\chi$, which we assume is negligible
compared to $m_\varphi$ (which is a natural assumption if $\chi$
stands for  a field of the Standard Model of particle physics or some
other possible particle from beyond the Standard model). Since
$m_\varphi \ll M_{\rm Pl}$ we have $q \gg 1$ for the values of $g$ of our interest. In this case, all modes
with $k < k_c$, where \cite{KLS}
\begin{equation}
 k_c \simeq  \frac{1}{2} \sqrt{g m_\varphi {\cal A}},
\label{kc}
\end{equation}
are excited exponentially and
\begin{equation}
\chi_k(z)  \propto  e^{\mu_k z},
\end{equation}
where $\mu_k$, called the {}Floquet exponent,  is of order one and it
depends only mildly on $k$ for the range of $k$ values mentioned
above. This is an example of broad parametric resonance.

The number density of $\chi$ particles produced during the preheating
grows as $\propto {\rm e}^{2 m_\varphi \mu t}$, where $t$ is the
duration of the preheating phase and $\mu$ is the maximum value of
$\mu_k$. The parametric resonance terminates due to the backreaction
of produced $\chi$ particles onto the homogeneous motion $\phi(t)$ of
$\varphi$, once the number density reaches the value \cite{KLS}
\begin{equation}
n_\chi \simeq \frac{m_\varphi^2 {\cal A}}{g}  ,
\end{equation}
where we have assumed that the resonance is fast and terminates before
the amplitude of the $\varphi$ oscillation decreases substantially due
to the expansion during preheating. Since the produced $\chi$
particles are non-relativistic at the end of preheating,
i.e.,~$m_{{\rm eff},\chi} \equiv g \vert \phi \vert \simeq g {\cal A}
> k_c$ (equivalent to $q>1$), the energy density of $\chi$ at the end
of preheating is evaluated to be
\begin{equation}
\rho_\chi \simeq g \vert\phi\vert n_\chi \simeq m_\varphi^2 {\cal A}^2 \; ,
\end{equation}
that is, most of the oscillation energy of $\varphi$ is transferred to
$\chi$ particles.

If, for example, one assumes a naive perturbative thermalization of $\chi$ through a
quartic coupling, e.g., with self-interacting Lagrangian density, 
\begin{equation} \label{IALag}
{\cal L}_\chi \, = \, - \frac{h}{4!} \, \chi^4 \, ,
\end{equation}
where $h$ is a dimensionless coupling constant, then
the interaction rate at the time of reheating would be \cite{Peskin:1995ev},
\begin{equation}
\Gamma^{\rm pert} \, = \, \langle n \sigma v \rangle \, \simeq \, 0.002 h^2 T_r^{\rm
  pert} \, ,
\end{equation}
where $n$ is the number density, $\sigma$ is the cross section of the interaction \eqref{IALag}, and $v$ is the mean relative velocity of the $\chi$ particles, which is taken to be unity in this case. Angular parentheses indicate spatial averaging, and $T_r^{\rm pert}$ is the reheating temperature in this treatment. Equating this to the Hubble parameter at temperature $T_r^{\rm pert}$ as the start of a radiation-dominate period,
we find 
\begin{equation}
T_r^{\rm pert} \, \simeq \, (2 \times 10^{16} \, {\rm GeV}) \,
g_*^{-1/2} h^2 \, , 
\end{equation}
where $g_*$ is the number of relativistic degrees of
freedom. 

On the other hand, if $\chi$ somehow decayed into relativistic
particles immediately after preheating and these particles thermalized
instantaneously, then the reheating temperature would be 
\begin{equation} T_r^{\rm inst} \simeq 1.3 g_*^{-1/4} \sqrt{m_\varphi {\cal A}} 
\simeq (2 \times 10^{15} \, {\rm GeV}) \, g^{-1/4} \, ,
\end{equation}
where $m_\varphi \simeq 5 \times 10^{-16} M_{\rm Pl}$ and 
${\cal A} \simeq 0.1 M_{\rm Pl}$ are
taken in the last step in view of the chaotic inflationary scenario.
These estimates of the reheating temperature, $T_r^{\rm pert}$ and
$T_r^{\rm inst}$, are based on too simplified assumptions and
should not be taken as the true reheating temperature without careful reasoning.
In the following, we perform a more rigorous analysis of the thermalization after
preheating of $\chi$ particles, with the use of the exact result of
the nonlinear Boltzmann equation obtained in Ref.~\cite{Denicol}. We
are here interested in the kinetic equilibration of the produced
$\chi$ particles and leave the study of the thermalization of the
residual energy in $\varphi$ to a separate study.

\section{The Boltzmann equation and its exact solution}
\label{sec3}

The fundamental object of our analysis is the phase space distribution
function $f({\bf x}, {\bf k}, t)$ of $\chi$ particles, where ${\bf k}$
are the comoving momentum vectors.  {}For homogeneous and isotropic
cosmology this function depends only on the magnitude $k$ of ${\bf
  k}$, i.e.,
\begin{equation}
f({\bf x}, {\bf k}, t)  \, =  \, f(k, t) \, \equiv \, f_k(t) \, .
\end{equation}
In terms of the distribution function, the number density of particles
is defined by
\begin{equation}
n(t) =  \frac{1}{2 \pi^2 a^3(t)} \int_0^{\infty} k^2 f_k(t) dk  ,
\end{equation}
and the energy density is
\begin{equation}
\rho(t)  =  \frac{1}{2 \pi^2 a^4(t)} \int_0^{\infty} k^3 f_k(t) dk  .
\end{equation}
In solving the Boltzmann equation below, we denote by $t_0$ the
initial time of the analysis (i.e., the time when the premises taken
in Ref.~\cite{Denicol} are met) and the number
density of $\chi$ particles at that time by $n_\chi$. We also introduce a
dimensionless time,
\begin{equation}
t'  \equiv  \frac{t}{l_0} \; ,
\end{equation}
by rescaling with respect to the initial mean free length $l_0$, which is given by
\begin{equation}
l_0  =  \frac{1}{\sigma_{\rm tot} n_\chi}  ,
\label{l0}
\end{equation}
where $\sigma_{\rm tot}$ is the total scattering cross-section and the
velocity of $\chi$ is taken to be unity under the assumption of
relativistic $\chi$ particles.  Here, we also follow
Ref.~\cite{Denicol} and assume that the total
cross-section $\sigma_{\rm tot}$ is independent of the momentum of
scattering particles for simplicity.  A more general cross-section
should not change this analysis qualitatively.

{}For the study of kinetic equilibration it is sufficient to study the
effects of $2 \rightarrow 2$ body interactions on the distribution
function.  Considering additional interactions could lead to a shorter
equilibration time, as we explicitly also show later on. The
relevant distribution function is obtained from the solution of the
Boltzmann equation, which in the case of a homogeneous and isotropic
distribution and written in the comoving frame, takes the
form~\cite{Denicol}
\begin{equation}
\frac{k}{a(t)} \frac{\partial}{\partial t} f_k  = {\cal C}_{\rm gain}
- {\cal C}_{\rm loss}  ,
\end{equation}
where the gain and loss terms are given by
\begin{eqnarray}
{\cal C}_{\rm gain}  &=& \frac{(2 \pi)^5}{2} \int_{k' p p'} s \sigma_T
\sqrt{-g} \delta^4(k + k' - p - p') f_p f_{p'}, \nonumber \\ {\cal
  C}_{\rm loss}  &=& \frac{(2 \pi)^5}{2} \int_{k' p p'} s \sigma_T
\sqrt{-g} \delta^4(k + k' - p - p') f_k f_{k'}, \nonumber \\
 \end{eqnarray}
where $s$ is the total energy and we define
\begin{equation}
\int_k \equiv \int \frac{d^3 k}{(2 \pi)^3 k^0 \sqrt{-\mathfrak{g}}} \; ,
\end{equation} 
with $\mathfrak{g}$ being the determinant of the metric.
The term ${\cal C}_{\rm gain}$ describes the particles produced by
scatterings, while ${\cal C}_{\rm loss}$ gives the loss of particles
as a consequence of such scatterings.

As discussed in Ref.~\cite{Denicol}, in the case of an initial
distribution of the form
\begin{equation} \label{indistrib}
f_k(0) \, = \, \lambda \frac{256}{243}  \frac{k}{T_0}  \exp\left( -
\frac{4 k}{3 T_0} \right) ,
\end{equation}
the Boltzmann equation in a FLRW spacetime has an exact solution,
\begin{equation} \label{evoldistrib}
f_k(\tau)  =  \frac{\lambda e^{- k / [K(\tau) T_0]}}{K^4(\tau)}
\left\{ 4 K(\tau) - 3 + \frac{k}{K(\tau) T_0} \left[1 - K(\tau)\right]
\right\}  ,
\end{equation}
where $\lambda$ is the fugacity, which we here set as one (i.e., we
assume negligible chemical potential for the system),  the new
dimensionless time variable $\tau$ in the above equation is defined as
\begin{equation}
\tau(t)  =  \int_{ t_0/l_0}^{ t/l_0}  dt' a^{-3}(t') ,
\label{tau}
\end{equation}
and the function $K(\tau)$ is
\begin{equation}
K(\tau)  =  1 - \frac{1}{4} e^{- \tau / 6}  .
\label{Ktau}
\end{equation}
This distribution approaches an equilibrium distribution on a time
scale given by $\tau \approx 6$.

The initial particle distribution given by Eq.~(\ref{indistrib}) is
qualitatively similar to the particle distribution produced by
broad parametric resonance with $T_0 \sim m_\varphi$. In both cases
the number density is nearly constant for all modes to the infrared of
a critical scale beyond which the number density falls off
exponentially when $k \gtrsim m_\varphi$. Hence, it is a good
approximation to take the distribution of $\chi$ particles once they become relativistic after
preheating to evolve according to Eq.~(\ref{evoldistrib}). In
particular, the distribution approaches a thermal one over a time
scale given by $\tau \sim {\cal O}(1)$. Different levels of agreement
with thermality can be parametrized by considering that
\begin{equation} \label{cond}
\tau  =  N ,
\end{equation}
where $N$ is some integer such that $N\leq 6$, for which $K(\tau)$ in 
Eq.~\eqref{Ktau} has not reached its static value $K \to 1$.

Let us now see how the above solution for the Boltzmann equation
applies to the present problem in this work.

\section{Application to kinetic equilibration following preheating}
\label{sec4}

In spite of the fact that the distribution of $\chi$ particles at the
end of preheating is not thermal,  the equation of state is
approximately that of a matter-dominated universe for some time after
preheating. The produced $\chi$ particles are also non-relativistic
right after the end of preheating, due to the mass modulated by the
coherent mode $\phi(t)$ of the inflaton, $m_{{\rm eff} , \chi}(t) = g
\phi(t)$. The $\chi$ particles are dominated by the modes around $k_c$
in Eq.~\eqref{kc} at the end of preheating, and from then on this
momentum redshifts due to expansion, $k(t) = k_c a(t_p) / a(t)$, where
$t_p$ denotes the time at the end of preheating. The effective mass
$m_{{\rm eff},\chi} \propto a^{-3/2}$ redshifts faster than momentum
$\propto a^{-1}$, and thus the produced $\chi$ particles eventually
becomes relativistic. This transition occurs at the moment when the
two quantities become equal, $m_{{\rm eff},\chi}(t_r) = k(t_r)$,
namely
\begin{equation}
\frac{a(t_r)}{a(t_p)} \simeq \frac{g^2 {\cal A}^2}{k_c^2}  ,
\label{time_p2r}
\end{equation}
where $t_r$ denotes the time when $\chi$ becomes relativistic and
${\cal A}$ is the amplitude of the $\varphi$ oscillation during
preheating as previously defined.  For  $t > t_r > t_p$, the
exact solution of the Boltzmann equation, Eq.~\eqref{evoldistrib}, is
valid.

As observed in \eqref{evoldistrib} with \eqref{tau}, the behavior in
approaching quasi-static distribution depends on the ratio $t_0/l_0$.
To evaluate $l_0$ by Eq.~\eqref{l0}, we consider as an example the quartic
self-interaction for the $\chi$ particles, as in Eq.~(\ref{IALag}). 
The total center-of-mass cross-section for scattering at
time $t=t_r$ is 
\begin{equation}
\sigma_{\rm tot} \, \simeq \, \sigma_0 \, \frac{48\pi^2 n_\chi(t_r)}{k^3(t_r)} \, ,
\end{equation}
where the cross-section in vacuum $\sigma_0 = h^2/(16 \pi k^2)$ \cite{Peskin:1995ev} 
is enhanced due to the presence of abundant $\chi$ particles~\cite{KLS}. Since the 
production of $\chi$ particles terminates when the produced modes start back-reacting 
to the producing $\varphi$ homogeneous modes, the number density of $\chi$ at this moment 
is \cite{KLS}
\begin{equation}
n_\chi(t_p) \, \simeq \, \frac{m_\varphi^2 {\cal A}}{g} \; .
\end{equation}
Thereafter, $n_\chi$ redshifts as $a^{-3}$ for $t_p < t <
t_r$.  Combining the above results, we find
\begin{equation}
l_0^{-1}  \simeq 3 \pi h^2 \frac{n_\chi^2(t_p)}{k_c^5} \,
\frac{a(t_p)}{a(t_r)} \simeq 3 \pi \, \frac{h^2}{g^4} \,
\frac{m_\varphi^4}{k_c^3}  .
\end{equation}
We now compare $l_0$ against the time scale of the expansion.
Identifying the initial time $t_0$ in Sec.~\ref{sec3} with $t_r$, 
the time when the solution \eqref{evoldistrib} begins to be valid, 
we have $t_0 = t_r = 1/[2 H(t_r)]$, where $H(t_r)$ is the Hubble rate at time $t_r$.
The energy density is roughly $\rho(t_p) \simeq m_\varphi^2 {\cal A}^2 /
2$ at the end of preheating, and it redshifts as $\rho(t) \propto
a^{-3}$ for $t > t_p$. Hence, we obtain that
\begin{equation}
t_0 \simeq \sqrt{\frac{3}{2}} \, \frac{M_{\rm Pl}}{m_\varphi {\cal A}}
\left[ \frac{a(t_r)}{a(t_p)} \right]^{3/2} \simeq \sqrt{\frac{3}{2}}
\, \frac{g^3 M_{\rm Pl} {\cal A}^2}{m_\varphi k_c^3}  .
\end{equation}
Therefore, the ratio for the initial scattering rate against the
expansion rate is
\begin{eqnarray}
\frac{t_0}{l_0} & \simeq & 2^{11/2} \cdot 3^{3/2} \pi \,
\frac{h^2}{g^4} \, \frac{M_{\rm Pl}}{{\cal A}} \nonumber \\  &\simeq &
7.4 \times 10^{15} \, h^2 \left( \frac{10^{-3}}{g} \right)^4 \left(
\frac{10^{-1} M_{\rm Pl}}{{\cal A}} \right)  .
\label{t0l0}
\end{eqnarray}
where Eq.~\eqref{kc} has been used for $k_c$.  The value for the
coupling constant $g$ needs to be $\lesssim 10^{-3}$ in order for
the quantum corrections coming from the interaction term \eqref{L_I} 
not to spoil the flatness of the inflaton potential. On
the other hand, we are interested in the regime of broad resonance $q
> 1$, which translates to $g > 2 m_\varphi / {\cal A}$. Thus, the
available range of $g$ is
\begin{equation}
10^{-4} \left( \frac{m_\varphi}{5 \times 10^{-6} \, M_{\rm Pl}}
\right) \left( \frac{10^{-1} \, M_{\rm Pl}}{{\cal A}} \right) < g
\lesssim 10^{-3}  .
\label{alpha_range}
\end{equation}
As for the quartic self-coupling $h$ for the $\chi$-particles, in
general, we can still have it satisfying $h \lesssim 8 \pi^2$ (which
is the typical numerical factor following the perturbative quantum
corrections involving the $\chi$ self-coupling) and still be in the
perturbative regime.  Thus, as seen in Eq.~\eqref{t0l0}, the time
scale of scattering due to the quartic self-interaction of $\chi$ is
much shorter than that of expansion at $t = t_0 \, (= t_r)$, $l_0 \ll t_0$, in
our scenario. Consequently, the condition given by Eq.~(\ref{cond})
becomes
\begin{equation} \label{result}
\delta t  \simeq  N l_0  ,
\end{equation}
where $\delta t$ is the time interval  from the beginning of kinetic 
equilibration to the time of its completion.

Our result Eq.~(\ref{result}) implies that kinetic equilibration
occurs on time scales much smaller than the expansion time scale,
i.e., almost instantaneously. This result does not agree with what is
obtained using simple perturbative methods and discussed  in
Sec.~\ref{sec:kinetic}.

{}From the above results, we can also give a prediction for the
duration of the whole process lasting from the end of inflation till
the completion of kinetic equilibration.  The onset of themalization
is at $t \simeq t_r$ in the scenario we have considered above. {}From
Eq.~\eqref{time_p2r}, the e-folding number from the end of inflation
to the onset time, $\Delta N_{pr}$, is then evaluated to be
\begin{eqnarray}
\Delta N_{pr} &=&  \ln \frac{a(t_r)}{a(t_p)} \nonumber \\ &\simeq & 4.38
+ \ln \frac{g}{10^{-3}} + \ln \frac{{\cal A}}{10^{-1} \, M_{\rm Pl}}
\nonumber \\ &&
- \ln \frac{m_{\varphi}}{5 \times 10^{-6} \, M_{\rm Pl}}  .
\label{efolds_p2r}
\end{eqnarray}
Since broad resonance is an efficient production mechanism for
particles and it terminates immediately after the inflaton oscillation
sets in, and since the thermalization completes almost
instantaneously, $\Delta N_{pr}$ well approximates the duration to
thermalize the produced $\chi$ particles after the end of inflation.

So far in this section, we have considered the thermalization of
$\chi$ particles after they become relativistic. Another possible
scenario is that $\chi$ could also decay into an already relativistic
particle, say $\sigma$, and $\sigma$ thermalizes to drive the
radiation-dominated epoch. Simplest relevant operators in such
processes are coupling terms of the form $-\kappa \chi \sigma^2$ for
the decay and $- \alpha \sigma^4 / 4!$ for the thermalization, where
$\kappa$ and $\alpha$ are the respective coupling constants. The
thermalization process by the $\sigma^4$ term parallels the one we
have discussed above, but the preheated $\chi$ field has to decay
first in such a scenario. The decay rate reads $\Gamma_{\chi \to 2
  \sigma} = \kappa^2 /(16\pi g \vert\phi\vert)$, and thus this decay
becomes effective when $\Gamma_{\chi \to 2 \sigma} \simeq H$, i.e.,
\begin{equation}
\frac{a(t)}{a(t_p)}  \simeq 2.7 \left( \frac{g}{\kappa^2} \,
\frac{m_\varphi {\cal A}^2}{M_{\rm Pl}} \right)^{1/3}  .
\label{time_p2sigma}
\end{equation}
This time scale comes in after the thermalization of $\chi$ by the
$\chi^4$ term if Eq.~\eqref{time_p2sigma} is larger than the result
given by Eq.~\eqref{time_p2r}, which gives us an upper bound for the value of the
coupling constant $\kappa$,
\begin{equation}
\kappa \lesssim 4.4 \times 10^{-8} \, M_{\rm Pl}  ,
\label{kappa}
\end{equation}
where we have used $g = 10^{-3}, \, m_\varphi = 5 \times 10^{-6}
M_{\rm Pl}$ and ${\cal A} = 0.1 M_{\rm Pl}$.  {}For  values of
$\kappa$ satisfying Eq.~(\ref{kappa}), the preheated $\chi$ particles
themselves thermalize quickly; otherwise, some intermediate particles
$\sigma$ may be involved in the thermalization process.  Hence, we can
see the result given by Eq.~\eqref{efolds_p2r} as giving an upper bound 
for the duration of the thermalization process. Any other
particle species taking part in this process can only render it more efficient. 

If the condition given by Eq.~\eqref{time_p2sigma} is satisfied and
thermalization completes through the $\chi^4$ interaction, the
reheating time is set by  Eqs.~\eqref{time_p2r} and
\eqref{efolds_p2r}. Since the energy density decreases as $\propto
a^{-3}$ during the period $t_p < t < t_r$, then using
\begin{eqnarray}
\rho(t_r) = \frac{\pi^2}{30} \, T_r^4
\simeq m_\varphi^2 {\cal A}^2 \left[ \frac{a(t_p)}{a(t_r)} \right]^3,
\end{eqnarray}
we have that the reheating temperature in this scenario is found to be
given by 
\begin{equation}
\begin{aligned}
T_r \simeq & \, 8.5 \times 10^{13} \, {\rm GeV} 
\\ & \times 
\left( \frac{10^{-3}}{g} \! \right)^{3/4}
\left( \frac{m_\varphi}{5 \times 10^{-6} M_{\rm Pl}} \right)^{5/4}
\left( \! \frac{0.1 M_{\rm Pl}}{{\cal A}} \right)^{1/4} \; .  
\end{aligned}
\end{equation}
This result should serve as the lower bound of the reheating
temperature in the scenario of preheating of the $\chi$ field after
chaotic inflation. As discussed above, thermalization may occur
earlier if $\chi$ decays into relativistic particles in an earlier
time when the $\chi$ particles are still non-relativistic.

\section{Conclusions and Discussion}
\label{conclusions}

Broad resonance preheating produces a state of $\chi$ particles which
have a similar initial distribution as a function of wavenumber as the
quanta in a relativistic heavy-ion collision  (RHIC) immediately after the
collision. We have used methods developed in Ref.~\cite{Denicol}
motivated by work on the thermalization after RHIC events to study the
kinetic equilibration of the $\chi$ quanta. Our analysis is based on
solving the Boltzmann equation for the phase space distribution of
$\chi$ quanta and the exact solution obtained in the case of a FLRW universe.

We find that the kinetic equilibration of the $\chi$ quanta after
preheating is fast on the Hubble time scale. We have not addressed
chemical equilibration which is required to obtain full thermal
equilibrium. Related to this, we have not studied the evolution of the
remnant energy in the inflaton field after preheating. It is possible
that one could set up coupled Boltzmann equations for the distribution
of both $\chi$ and $\phi$ particles, taking into account the
interactions, but the calculation would be more involved. However, our
results already show that thermalization after preheating may be much more
rapid than what is concluded from leading-order perturbative
computations. This leads to a higher reheating temperature.
{}Furthermore, our results have allowed us to make a clear prediction for
the total number of e-folds, following the end of inflation, required
to achieve thermalization in the kinetic regime.  The total time
duration of the reheating phase is important for precise comparisons
of the predictions of inflationary models with cosmological
observations (see, e.g., Ref.~\cite{Dai:2014jja}). Our results may
also be important for physical phenomena which take place early in the
radiation phase of Standard Big Bang cosmology, e.g.~baryogenesis.

The results obtained in Sec.~\ref{sec4} can also be applied to
determine the entropy in the $\chi$ field. The entropy density
associated with the phase space distribution $f_k$ is~\cite{Denicol}
\begin{equation}
s \, \equiv \, \int_k k f_k \bigl( {\rm ln} f_k - 1 \bigr) \; .
\end{equation}
Since $f_k$ approaches its kinetic equilibrium distribution on the
time scale $\tau = N$, we conclude that the entropy density in $\chi$
fluctuations approaches its equilibrium value on the same time
scale, and the resulting density should read
\begin{equation}
s \, \sim \, T_R^3 \, .
\end{equation}
This is particularly important when setting for instance the abundance
of possible dark matter candidates. Our results then show that these
abundances can in principle be precisely defined soon after the
kinetic equilibration of the produced particles from preheating. 

In this work we have studied kinetic equilibration in the context of
models in which inflation terminates in a phase of broad band
parametric resonance.  However, our analysis is equally applicable to
models with tachyonic preheating~\cite{tachyonic}, since in both cases
the fluctuation  modes with wavenumbers below a critical value $k_c$
are exactly the ones that are excited,  whereas shorter wavelength
modes are not. Thus, the procedure used here applies similarly to the
case of tachyonic preheating.

\section*{Acknowledgement}
\noindent

The research at McGill is supported in part by funds from NSERC and
from the  Canada Research Chair program. R.O.R is partially supported
by research grants from Conselho Nacional de Desenvolvimento
Cient\'{\i}fico e Tecnol\'ogico (CNPq), grant No. 302545/2017-4 and
Funda\c{c}\~ao Carlos Chagas Filho de Amparo \`a Pesquisa do Estado do
Rio de Janeiro (FAPERJ), grant No.  E - 26/202.892/2017.



\begin{thebibliography}{99}

\bibitem{Guth} A. H. Guth,  ``The Inflationary Universe: A Possible
  Solution To The Horizon And Flatness Problems,''  Phys.\ Rev.\  D
  {\bf 23}, 347 (1981);\\
R.~Brout, F.~Englert and E.~Gunzig,
 ``The Creation Of The Universe As A Quantum Phenomenon,
 '' Annals Phys.\  {\bf 115}, 78 (1978);\\
K. Sato,
 ``First Order Phase Transition of A Vacuum And Expansion Of The Universe,
 '' Mon.\ Not.\ Roy.\ Astron.\ Soc.\  {\bf 195}, 467 (1981);\\
L.~Z.~Fang, 
``Entropy Generation in the Early Universe by Dissipative Processes Near the Higgs' Phase Transitions,
'' Phys.\ Lett.\ B {\bf 95}, 154 (1980);\\
  A.~A.~Starobinsky,
  ``Spectrum of relict gravitational radiation and the early state of the universe,''
  JETP Lett.\  {\bf 30}, 682 (1979)
  [Pisma Zh.\ Eksp.\ Teor.\ Fiz.\  {\bf 30}, 719 (1979)];
\\
A.~A.~Starobinsky,
 ``A New Type Of Isotropic Cosmological Models Without Singularity,'' Phys.\ Lett.\ B {\bf 91}, 99 (1980).

\bibitem{Berera:2008ar} 
  A.~Berera, I.~G.~Moss and R.~O.~Ramos,
  ``Warm Inflation and its Microphysical Basis,''
  Rept.\ Prog.\ Phys.\  {\bf 72}, 026901 (2009)
  doi:10.1088/0034-4885/72/2/026901
  [arXiv:0808.1855 [hep-ph]].
  
\bibitem{DL}
A.~D.~Dolgov and A.~D.~Linde,
  ``Baryon Asymmetry in Inflationary Universe,''
  Phys.\ Lett.\  {\bf 116B}, 329 (1982).
  doi:10.1016/0370-2693(82)90292-1
  
\bibitem{AFW}
L.~F.~Abbott, E.~Farhi and M.~B.~Wise,
  ``Particle Production in the New Inflationary Cosmology,''
  Phys.\ Lett.\  {\bf 117B}, 29 (1982).
  doi:10.1016/0370-2693(82)90867-X
  
\bibitem{TB}
J.~H.~Traschen and R.~H.~Brandenberger,
  ``Particle Production During Out-of-equilibrium Phase Transitions,''
  Phys.\ Rev.\ D {\bf 42}, 2491 (1990).
  doi:10.1103/PhysRevD.42.2491
  
\bibitem{DK}
A.~D.~Dolgov and D.~P.~Kirilova,
  ``On Particle Creation By A Time Dependent Scalar Field,''
  Sov.\ J.\ Nucl.\ Phys.\  {\bf 51}, 172 (1990)
  [Yad.\ Fiz.\  {\bf 51}, 273 (1990)].
  
\bibitem{Floquet}
McLachlan, N. W. ``Theory and applications of Mathieu equations.'' Clarendon, Oxford (1947).
  
\bibitem{KLS}
L.~Kofman, A.~D.~Linde and A.~A.~Starobinsky,
  ``Reheating after inflation,''
  Phys.\ Rev.\ Lett.\  {\bf 73}, 3195 (1994)
  doi:10.1103/PhysRevLett.73.3195
  [hep-th/9405187].
  
\bibitem{STB}
Y.~Shtanov, J.~H.~Traschen and R.~H.~Brandenberger,
  ``Universe reheating after inflation,''
  Phys.\ Rev.\ D {\bf 51}, 5438 (1995)
  doi:10.1103/PhysRevD.51.5438
  [hep-ph/9407247].
  
\bibitem{KLS2}
L.~Kofman, A.~D.~Linde and A.~A.~Starobinsky,
  ``Towards the theory of reheating after inflation,''
  Phys.\ Rev.\ D {\bf 56}, 3258 (1997)
  doi:10.1103/PhysRevD.56.3258
  [hep-ph/9704452].
  
\bibitem{ABCM}
R.~Allahverdi, R.~Brandenberger, F.~Y.~Cyr-Racine and A.~Mazumdar,
  ``Reheating in Inflationary Cosmology: Theory and Applications,''
  Ann.\ Rev.\ Nucl.\ Part.\ Sci.\  {\bf 60}, 27 (2010)
  doi:10.1146/annurev.nucl.012809.104511
  [arXiv:1001.2600 [hep-th]].
  
\bibitem{Karouby}  
M.~A.~Amin, M.~P.~Hertzberg, D.~I.~Kaiser and J.~Karouby,
  ``Nonperturbative Dynamics Of Reheating After Inflation: A Review,''
  Int.\ J.\ Mod.\ Phys.\ D {\bf 24}, 1530003 (2014)
  doi:10.1142/S0218271815300037
  [arXiv:1410.3808 [hep-ph]].

\bibitem{original}
J.~R.~Ellis, K.~Enqvist, D.~V.~Nanopoulos and K.~A.~Olive,
  ``Inflationary Fluctuations, Entropy Generation and Baryogenesis,''
  Phys.\ Lett.\ B {\bf 191}, 343 (1987).
  doi:10.1016/0370-2693(87)90620-4;\\
S.~Davidson and S.~Sarkar,
  ``Thermalization after inflation,''
  JHEP {\bf 0011}, 012 (2000)
  doi:10.1088/1126-6708/2000/11/012
  [hep-ph/0009078];\\
  R.~Allahverdi,
  ``Thermalization after inflation and reheating temperature,''
  Phys.\ Rev.\ D {\bf 62}, 063509 (2000)
  doi:10.1103/PhysRevD.62.063509
  [hep-ph/0004035];\\
  R.~Allahverdi and M.~Drees,
  ``Thermalization after inflation and production of massive stable particles,''
  Phys.\ Rev.\ D {\bf 66}, 063513 (2002)
  doi:10.1103/PhysRevD.66.063513
  [hep-ph/0205246];\\
  P.~Jaikumar and A.~Mazumdar,
  ``Postinflationary thermalization and hadronization: QCD based approach,''
  Nucl.\ Phys.\ B {\bf 683}, 264 (2004)
  doi:10.1016/j.nuclphysb.2004.02.015
  [hep-ph/0212265];\\
  D.~I.~Podolsky, G.~N.~Felder, L.~Kofman and M.~Peloso,
  ``Equation of state and beginning of thermalization after preheating,''
  Phys.\ Rev.\ D {\bf 73}, 023501 (2006)
  doi:10.1103/PhysRevD.73.023501
  [hep-ph/0507096];
\\
  R.~Allahverdi and A.~Mazumdar,
  ``Supersymmetric thermalization and quasi-thermal universe: Consequences for gravitinos and leptogenesis,''
  JCAP {\bf 0610}, 008 (2006)
  doi:10.1088/1475-7516/2006/10/008
  [hep-ph/0512227];\\
  R.~Allahverdi and A.~Mazumdar,
  ``Reheating in supersymmetric high scale inflation,''
  Phys.\ Rev.\ D {\bf 76}, 103526 (2007)
  doi:10.1103/PhysRevD.76.103526
  [hep-ph/0603244];\\
  A.~Mazumdar and B.~Zaldivar,
  ``Quantifying the reheating temperature of the universe,''
  Nucl.\ Phys.\ B {\bf 886}, 312 (2014)
  doi:10.1016/j.nuclphysb.2014.07.001
  [arXiv:1310.5143 [hep-ph]].
  
\bibitem{numerical}
R.~Micha and I.~I.~Tkachev,
  ``Relativistic turbulence: A Long way from preheating to equilibrium,''
  Phys.\ Rev.\ Lett.\  {\bf 90}, 121301 (2003)
  doi:10.1103/PhysRevLett.90.121301
  [hep-ph/0210202];\\
  R.~Micha and I.~I.~Tkachev,
  ``Turbulent thermalization,''
  Phys.\ Rev.\ D {\bf 70}, 043538 (2004)
  doi:10.1103/PhysRevD.70.043538
  [hep-ph/0403101].
  
\bibitem{Berges}  
J.~Berges and J.~Serreau,
  ``Parametric resonance in quantum field theory,''
  Phys.\ Rev.\ Lett.\  {\bf 91}, 111601 (2003)
  doi:10.1103/PhysRevLett.91.111601
  [hep-ph/0208070];\\
J.~Berges, D.~Gelfand and J.~Pruschke,
  ``Quantum theory of fermion production after inflation,''
  Phys.\ Rev.\ Lett.\  {\bf 107}, 061301 (2011)
  doi:10.1103/PhysRevLett.107.061301
  [arXiv:1012.4632 [hep-ph]];\\
J.~Berges, S.~Borsanyi and J.~Serreau,
  ``Thermalization of fermionic quantum fields,''
  Nucl.\ Phys.\ B {\bf 660}, 51 (2003)
  doi:10.1016/S0550-3213(03)00261-X
  [hep-ph/0212404].
    
\bibitem{RHIC}
J.~Berges, S.~Borsanyi and C.~Wetterich,
  ``Prethermalization,''
  Phys.\ Rev.\ Lett.\  {\bf 93}, 142002 (2004)
  doi:10.1103/PhysRevLett.93.142002
  [hep-ph/0403234];\\
J.~Berges, K.~Boguslavski, S.~Schlichting and R.~Venugopalan,
  ``Turbulent thermalization process in heavy-ion collisions at ultrarelativistic energies,''
  Phys.\ Rev.\ D {\bf 89}, no. 7, 074011 (2014)
  doi:10.1103/PhysRevD.89.074011
  [arXiv:1303.5650 [hep-ph]].
  
\bibitem{Denicol}
D.~Bazow, G.~S.~Denicol, U.~Heinz, M.~Martinez and J.~Noronha,
  ``Nonlinear dynamics from the relativistic Boltzmann equation in the Friedmann-Lemaître-Robertson-Walker spacetime,''
  Phys.\ Rev.\ D {\bf 94}, no. 12, 125006 (2016)
  doi:10.1103/PhysRevD.94.125006
  [arXiv:1607.05245 [hep-ph]].

\bibitem{Linde}
A.~D.~Linde,
  ``Chaotic Inflation,''
  Phys.\ Lett.\  {\bf 129B}, 177 (1983).
  doi:10.1016/0370-2693(83)90837-7
  
\bibitem{MFB}
V.F. Mukhanov, H.A. Feldman and R.H. Brandenberger, 
``Theory of Cosmological Perturbations'' 
Physics Reports \textbf{215}, 203 (1992).

\bibitem{RHBfluctRev}
 R. H. Brandenberger,
     ``Lectures on the theory of cosmological perturbations'' 
     Lect. Notes Phys. \textbf{646}, 127 (2004) 
     {[}arXiv:hep-th/0306071{]}.
  
\bibitem{Peskin:1995ev} 
  M.~E.~Peskin and D.~V.~Schroeder,
  ``An Introduction to quantum field theory,''
  (Addison-Wesley, NY, 1995).
  
\bibitem{Dai:2014jja} 
  L.~Dai, M.~Kamionkowski and J.~Wang,
  ``Reheating constraints to inflationary models,''
  Phys.\ Rev.\ Lett.\  {\bf 113}, 041302 (2014)
  doi:10.1103/PhysRevLett.113.041302
  [arXiv:1404.6704 [astro-ph.CO]].

\bibitem{tachyonic} 
G.~N.~Felder, J.~Garcia-Bellido, P.~B.~Greene, L.~Kofman, A.~D.~Linde and I.~Tkachev,
  ``Dynamics of symmetry breaking and tachyonic preheating,''
  Phys.\ Rev.\ Lett.\  {\bf 87}, 011601 (2001)
  doi:10.1103/PhysRevLett.87.011601
  [hep-ph/0012142];\\
E.~J.~Copeland, S.~Pascoli and A.~Rajantie,
  ``Dynamics of tachyonic preheating after hybrid inflation,''
  Phys.\ Rev.\ D {\bf 65}, 103517 (2002)
  doi:10.1103/PhysRevD.65.103517
  [hep-ph/0202031].

\end{thebibliography}
\end{document}